\documentclass[preprint,showpacs,prl]{revtex4}
\usepackage{amsfonts}
\usepackage{amsmath}
\usepackage{amssymb}
\usepackage{graphicx}

\input{tcilatex}

\begin{document}

\title[Short title for running header]{NMR experimental realization of
seventh-order coupling transformations and the seven-qubit
modified Deutsch-Jozsa algorithm }
\author{Daxiu Wei, Jun Luo, Xiaodong Yang, Xianping Sun, Xizhi Zeng, Maili
Liu, Shangwu Ding, and Mingsheng Zhan} \affiliation{State Key
Laboratory of Magnetic Resonance and Atomic and Molecular
Physics,Wuhan Institute of Physics and Mathematics, Chinese
Academy of Sciences, Wuhan 430071, People's Republic of China}

\begin{abstract}
We propose a scalable method on the basis of nth-order coupling
operators to construct f-dependent phase transformations in the
n-qubit modified Deutsch-Jozsa (D-J) quantum algorithm. The novel
n-qubit entangling transformations are easily implemented via
J-couplings between neighboring spins. The seven-qubit modified
D-J quantum algorithm and seventh-order coupling transformations
are then experimentally demonstrated with liquid state nuclear
magnetic resonance (NMR) techniques. The method may offer the
possibility of creating generally entangled states of n qubits and
simulating n-body interactions on n-qubit NMR quantum computers.
\end{abstract}

\maketitle

By using the characteristics of quantum mechanics, quantum computers (QC)
are faster than their classical counterparts when performing certain
computational tasks such as factorizing a large number \cite{shor},
searching unsorted database \cite{grover}, and especially when simulating
quantum systems themselves \cite{lloyd}. Among all quantum algorithms,
Deutsch's algorithm \cite{deutsch,cleve} was not only the first to be
proposed, but also the first to be demonstrated \cite{jcfm} using liquid
state nuclear magnetic resonance (NMR) techniques \cite{cory,nag}. An
experimental realization of Deutsch-Jozsa (D-J) algorithm has not only
demonstrated that QCs have an information-processing capability much greater
than that of their classical counterparts, but also provided crucial insight
into more impressive quantum algorithms \cite{man}. Based on the Cleve
version \cite{cleve}, two-, three- or five-qubit D-J algorithm has been
implemented with NMR spectroscopy \cite{jcfm}. All of the previous D-J
algorithms required n+1 qubit quantum systems to realize the n-bit D-J
algorithm. However, Collins \textit{et al}. \cite{col1a} proposed a modified
version of the D-J algorithm, which allowed an implementation of the n-qubit
D-J algorithm by using n qubits. Up to now, the modified D-J algorithm has
been implemented only on three-qubit spin systems \cite{col1a}.

Extending qubit number in quantum computation has been a subject of much
recent interest \cite{bc}. Currently only NMR technique is used to
experimentally demonstrate quantum information processing with more than
four qubits. Such progresses \cite{jones0} include seven-qubit ``cat''-state
\cite{knill} and Shor's algorithm \cite{lieven}, five-qubit D-J algorithm
\cite{jcfm}, five-qubit order finding algorithm \cite{lieve} and five-qubit
error correction benchmark \cite{knill00}. However, experimentally it is a
technical challenge to extend to more qubits because of the low
signal-to-noise ratio (SNR) in NMR experiments and furthermore, the SNR's
exponential decrease with the increase of qubit number \cite{warren}. In the
way to get more qubits, there are two obstacles. First, although efficient
methods have been developed to refocus the interaction of single-spin and
two spins \cite{jones1}, it is not easy to control the coherent evolution
between two specific spins. This is because the coupling network becomes
more complex when more qubits are involved. Second, it is difficult to find
any molecule that has appreciable J-coupling configuration between any pair
of spins. Fortunately, M\'{a}di \textit{et al}. \cite{madi} have used
quantum swap gates to surmount the second difficulty.

In this letter, we propose a simple method to construct n-qubit entangling
transformations on the basis of nth-order coupling operators. The method we
present is general and scalable. The key to the method is only to use the
J-couplings between neighbouring spins to implement nth-order coupling
transformations. With this method, we can construct n-qubit entangling
transformations for balanced functions in the modified D-J algorithm. And
then we experimentally demonstrate the seven-qubit modified D-J algorithm
and seventh-order coupling transformations using NMR.

\textit{The modified D-J algorithm}.---A D-J problem \cite{deutsch,cleve}
is: given Boolean function f(x): $\{0,1\}^{n}\longrightarrow \{0,1\}$, how
many evaluations are necessary to determine f(x) being constant or
balanced---constant functions f(x) always give the same output (all 0 or all
1) for all input values, but balanced functions f(x) have an equal number of
0 outputs as 1s. In a modified version of the D-J algorithm \cite{col1a},
the key to the scheme is to perform an f-controlled gate \cite{col1a}, whose
transformation on the basis elements $|x>\equiv |x_{n-1}\cdot \cdot \cdot
x_{0}>$ [$x_{i}\in (0,1)$] is defined as \cite{col1a}:

\begin{equation}
U_{f}|x>:=(-1)^{f(x)}|x>.  \label{ufq}
\end{equation}
\newline
The unitary transformation $U_{f}$ is an f-dependent phase operation on 2$%
^{n}$ eigenstates of an n-qubit quantum computer. For example, the number of
functions for the seven-bit D-J problem is $\binom{128}{64}$+2, which is a
large number. In general, it is difficult to express the unitary
transformation $U_{f}$ as the product of operators allowed by the Ising-type
Hamiltonian of spin-1/2 systems. Fortunately, we can classify the $U_{f}$ s
as seven representative types of unitary transformations namely,
non-entangling, and from two-qubit entangling to seven-qubit entangling.

\textit{Constructing n-qubit f-dependent phase transformations via nth-order
coupling operators}.---We consider 9 explicit transformations, as shown in
table 1. The transformations ($U_{f4}-U_{f9}$) in table 1 correspond to the
two-qubit to seven-qubit entangling operators, respectively. From the
transformations ($U_{f4}-U_{f9}$), we can easily extend the entangling
transformations to an n-qubit case. For example, an f-dependent phase
transformation for any balanced function in the n-bit D-J algorithm can read:

\begin{eqnarray}
U_{n-bit-f} &=&\frac{1}{2}\left( E^{(1)}\otimes \cdot \cdot \cdot \otimes
E^{(n-2)}\otimes \sigma _{z}^{(n-1)}\otimes E^{(n)}+E^{(1)}\otimes \cdot
\cdot \cdot \otimes E^{(n-2)}\otimes \sigma _{z}^{(n-1)}\otimes \sigma
_{z}^{(n)}\right)  \notag \\
&&+\frac{1}{2}\left( \sigma _{z}^{(1)}\otimes \cdot \cdot \cdot \otimes
\sigma _{z}^{(n-1)}\otimes E^{(n)}-\sigma _{z}^{(1)}\otimes \cdot \cdot
\cdot \otimes \sigma _{z}^{(n)}\right) ,  \label{fphase}
\end{eqnarray}
\newline
where $\sigma _{z}$ is the Pauli matrix, $E$ is a 2$\times 2$ unit matrix,
and superscripts label the qubits involved. The transformation $U_{n-bit-f}$
is diagonal in the 2$^{n}$ eigenbasis vectors. It is interesting to note
that the transformation $U_{n-bit-f}$ cannot be decomposed as direct
products of single-spin operators and is hence an entangling operation. The
n-qubit entangling transformation $U_{n-bit-f}$ in the modified D-J
algorithm can be constructed on the basis of second, nth, and (n-1)th-order
coupling operators. In order to realize the n-qubit entangling
transformation $U_{n-bit-f}$ experimentally, $U_{n-bit-f}$ can be decomposed
with the product operator basis set:

\begin{align}
U_{n-bit-f}& =\exp [i\pi ]\left\{ \exp \left[ -i\frac{\pi }{2}\sigma
_{z}^{(1)}\cdot \cdot \cdot \sigma _{z}^{(n-1)}\right] \otimes
E^{(n)}\right\} \exp \left[ i\frac{\pi }{2}\sigma _{z}^{(1)}\cdot \cdot
\cdot \sigma _{z}^{(n)}\right]  \label{nbit} \\
& \left\{ E^{(1)}\otimes \cdot \cdot \cdot \otimes E^{(n-2)}\otimes \exp
\left[ -i\frac{\pi }{2}\sigma _{z}^{(n-1)}\sigma _{z}^{(n)}\right] \right\}
\notag \\
& \left\{ E^{(1)}\otimes \cdot \cdot \cdot \otimes E^{(n-2)}\otimes \exp
\left[ -i\frac{\pi }{2}\sigma _{z}^{(n-1)}\right] \otimes E^{(n)}\right\} .
\notag
\end{align}
\newline
Following a procedure similar to that given in Ref. \cite{kim}, we perform
the nth-order coupling operation $\exp \left[ i\frac{\pi }{2}\sigma
_{z}^{(1)}\cdot \cdot \cdot \sigma _{z}^{(n)}\right] $ in terms of the
operators allowed by the Hamiltonian of spin-1/2 systems, i.e., with radio
frequency pulses and scalar J-couplings between two neighboring spins. For
example, a quantum circuit for performing a seven-qubit entangling
transformation ($U_{f9}$) is demonstrated in Fig. 1(a).

\textit{Experimental realization}.---A quantum circuit shown in Fig. 1(b)
can be used to implement the seven-qubit modified D-J algorithm. This
circuit begins with an initial state (\TEXTsymbol{\vert}0000000\TEXTsymbol{>}%
). In fact, however, the statistical mixture of pure states can be used \cite%
{jcfm,col1a,zhou} as input states, since the D-J algorithm can accept the
thermal equilibrium state as an input \cite{jcfm,col1a,zhou}. There are two
important operations in Fig. 1(b), namely pseudo-Hadamard and f-dependant
phase gates. $h$ and $h^{-1}$ are pseudo-Hadamard gates, which can be
implemented by 90 degree pulses along $\pm $y axes, respectively. The key to
this approach is to realize seven-qubit f-dependant phase transformations
shown in table 1 and Fig. 1(a).

In order to read out results, it is necessary to apply a selective ($\pi $/2)%
$_{y}$ pulse to the corresponding spin. According to phases (absorption or
emission) of the $^{13}$C and $^{1}$H NMR signals, NMR spectra clearly
indicate where the system is. Therefore we can determine the constant or
balanced nature of the function (Fig. 1(b)). Fortunately, we can cancel the
final pseudo-Hadmard transformations by the ($\pi $/2)$_{y}$ read-out pulse
used in NMR experiments. In order to improve experimental results, we have
used the first-order phase correction methods \cite{lmkv}, and acquired NMR
spectra with 96 scans.

We have selected a seven-qubit spin system [U-$^{13}$C$_{4}$-labeled
crotonic acid ($^{13}$C$^{1}$H$_{3}^{3}$ $^{13}$C$^{2}$H$^{1}$=$^{13}$C$^{3}$%
H$^{2}$ $^{13}$C$^{4}$O$_{2}$H, Cambridge Isotope Laboratories Inc. Cat. No.
CLM-6118), where right superscripts label the qubits] \cite{knill}. The
sample of 20mg crotonic acid was dissolved in deuterated acetone, degassed
and flame-sealed in a standard 5mm NMR test tube, the coupling constants of
which can be found in Ref. \cite{knill}. We began the experiment from a
thermal equilibrium state. We have used four carbons and three protons as
the seven qubits. We carried out the experiment with a Varian INOVA 600 NMR
spectrometer. All NMR experiments were conducted at 25 $^{0}$C. The shape of
the soft pulse was Gaussian. The J-coupling interactions in NMR spin systems
have been used to implement entangling transformations required for the
seven-qubit modified D-J algorithm.

\textit{Results and discussion}.---Typical experimental results are shown in
figures 2 and 3. The experimental spectra in figure 2 correspond to the
constant function ($U_{f1}$), which is directly obtained after applying a 90
degree pulse to the thermal equilibrium state. The spectra in Fig. 2 served
as reference spectra. The phase of the reference spectra was adjusted so
that signals from all the seven spins appear in absorption (positive phase).
The experimental spectra in Fig. 3 correspond to the balanced functions ($%
U_{f2}$, $U_{f4}$, $U_{f5}$, and $U_{f9}$), respectively. Compared with the
reference spectra in Fig. 2, the spectra in Fig. 3 show that there is at
least one line with a $\pi $ phase difference (in emission). By determining
the relative phase of the signals from the seven spins, we can determine
that the functions corresponding to Fig. 3 are balanced. It should be
noticed that only a single function call is used.

Fig. 2 shows that the spectra of C$^{1}$ contain 128 peaks. This indicates
that spin C$^{1}$ interacts with 7 neighbors. This results from the
J-couplings of the three other $^{13}$C, the two protons (H$^{1}$, H$^{2}$)
and the three methyl protons (H$^{3}$). It should be noted that the
J-coupling effect of the methyl protons on the spin C$^{1}$ is the same as
that of two protons.

As for the thermal equilibrium state used as an input state in our
experiments, the investigations \cite{jcfm,col1a,zhou} have shown that the
D-J algorithm can be implemented by starting with thermal rather than pure
initial states. The reason is that the thermal equilibrium state has similar
effect as the effective pure state.

There are imperfect phases in the experimental results. The signals do not
exhibit pure absorption or pure emission lineshapes. The phase errors mainly
come from (a) the imperfection of the selective pulses, (b) the inaccuracy
of the 90 and 180 degree pulses, (c) inaccurate refocusing of chemical
shifts during the J-coupling delays, and (d) J-coupling evolution during
long selective pulses.

It should be noted that n-qubit entangling transformations in the modified
D-J algorithm can be performed with radio frequency pulses and scalar
J-couplings between two neighboring spins. The scheme may be used to create
generally entangled states with n qubits on the basis of the interactions
between neighboring spins. This is very important for demonstrating the role
of quantum entanglement in quantum information processing.

We used seventh-order coupling operators to construct a seven-qubit
entangling transformation. We have successfully realized the seven-qubit
entangling transformation in the context of the modified D-J algorithm.
Therefore we have experimentally performed the seventh-order coupling
operator. In fact, there is no seventh-order coupling in liquid state NMR
systems. Our experimental results show that seven-body interactions can be
simulated with such an experimental method. This will be useful in
simulating many-body interactions in other quantum systems.

The NMR experimental realization of multi-qubit quantum algorithms is a
difficult proposition. In a multi-qubit NMR QC, it is necessary to suppress
the interactions coming from the other qubits when realizing the quantum
logic gate between two given spins. Our experimental results have
demonstrated that the control of multi-qubit's evolution is successful.

\textit{Conclusions}.---On the basis of constructing f-dependent phase
transformations in the modified D-J quantum algorithm with nth-order
coupling operators, we have experimentally tested the modified D-J algorithm
for seven qubits and performed seventh-order coupling transformations. This
may open a way to experimentally simulate n-body interactions and to realize
n-qubit entangling transformations on NMR quantum computers.

\newpage

Table 1. The transformations \textit{U}$_{f}$ and their defintions. $%
E^{(A,B,C,\cdot \cdot \cdot )}$ and $\sigma _{z}^{(A,B,C,\cdot \cdot \cdot
)} $ mean $E^{(A)}\otimes E^{(B)}\otimes E^{(C)}\otimes \cdot \cdot \cdot $,
and $\sigma _{z}^{(A)}\otimes \sigma _{z}^{(B)}\otimes \sigma
_{z}^{(C)}\otimes \cdot \cdot \cdot $, respectively. C$^{i}$ (i=1,2,3, and
4) and H$^{j}$ (j=1,2, and 3) are carbon and hydrogen nuclei used in
experiments.\bigskip\

\begin{tabular}{cc}
\hline\hline
\textit{U}$_{f}$ & Defintions \\ \hline\hline
Constant: &  \\
\textit{U}$_{f1}$ & $E^{(C^{1},C^{2},C^{3},C^{4},H^{1},H^{2},H^{3})}$ \\
Balanced: &  \\
\textit{U}$_{f2}$ & $\sigma _{z}^{(C^{2},C^{3})}\otimes
E^{(C^{1},C^{4},H^{1},H^{2},H^{3})}$ \\
\textit{U}$_{f3}$ & $\sigma
_{z}^{(C^{1},C^{2},C^{3},C^{4},H^{1},H^{2},H^{3})}$ \\
\textit{U}$_{f4}$ & $\frac{1}{2}\left( E^{(C^{3},C^{4})}+\sigma
_{z}^{(C^{3})}\otimes E^{(C^{4})}+E^{(C^{3})}\otimes \sigma
_{z}^{(C^{4})}-\sigma _{z}^{(C^{3},C^{4})}\right) \otimes \sigma
_{z}^{(C^{2})}\otimes E^{(C^{1},H^{1},H^{2},H^{3})}$ \\
\begin{tabular}{l}
\textit{U}$_{f5}$%
\end{tabular}
&
\begin{tabular}{l}
$\frac{1}{2}\left( E^{(C^{2})}\otimes \sigma _{z}^{(C^{1})}\otimes
E^{(H^{3})}+E^{(C^{2})}\otimes \sigma _{z}^{(C^{1},H^{3})}+\sigma
_{z}^{(C^{2},C^{1})}\otimes E^{(H^{3})}-\sigma
_{z}^{(C^{2},C^{1},H^{3})}\right) $ \\
$\ \ \ \ \ \ \ \ \ \otimes E^{(C^{3},C^{4},H^{1},H^{2})}$%
\end{tabular}
\\
\begin{tabular}{l}
\textit{U}$_{f6}$%
\end{tabular}
&
\begin{tabular}{l}
$\frac{1}{2}\left( E^{(C^{1},C^{2})}\otimes \sigma _{z}^{(C^{3})}\otimes
E^{(C^{4})}+E^{(C^{1},C^{2})}\otimes \sigma _{z}^{(C^{3},C^{4})}\right)
\otimes E^{(H^{1},H^{2},H^{3})}$ \\
+$\frac{1}{2}\left( \sigma _{z}^{(C^{1},C^{2},C^{3})}\otimes
E^{(C^{4})}-\sigma _{z}^{(C^{1},C^{2},C^{3},C^{4})}\right) \otimes
E^{(H^{1},H^{2},H^{3})}$%
\end{tabular}
\\
\begin{tabular}{l}
\textit{U}$_{f7}$%
\end{tabular}
&
\begin{tabular}{l}
$\frac{1}{2}\left( E^{(C^{1},C^{2},C^{3})}\otimes \sigma
_{z}^{(C^{3})}\otimes E^{(H^{1})}+E^{(C^{1},C^{2},C^{3})}\otimes \sigma
_{z}^{(C^{4})}\otimes \sigma _{z}^{(H^{1})}\right) \otimes E^{(H^{2},H^{3})}$
\\
+$\frac{1}{2}\left( \sigma _{z}^{(C^{1},C^{2},C^{3},C^{4})}\otimes
E^{(H^{1})}-\sigma _{z}^{(C^{1},C^{2},C^{3},C^{4},H^{1})}\right) \otimes
E^{(H^{2},H^{3})}$%
\end{tabular}
\\
\begin{tabular}{l}
\textit{U}$_{f8}$%
\end{tabular}
&
\begin{tabular}{l}
$\frac{1}{2}\left( E^{(C^{1},C^{2},C^{3},C^{4})}\otimes \sigma
_{z}^{(H^{2})}\otimes E^{(H^{2})}+E^{(C^{1},C^{2},C^{3},C^{4})}\otimes
\sigma _{z}^{(H^{1},H^{2})}\right) \otimes E^{(H^{3})}$ \\
+$\frac{1}{2}\left( \sigma _{z}^{(C^{1},C^{2},C^{3},C^{4},H^{1})}\otimes
E^{(H^{2})}-\sigma _{z}^{(C^{1},C^{2},C^{3},C^{4},H^{1},H^{2})}\right)
\otimes E^{(H^{3})}$%
\end{tabular}
\\
\begin{tabular}{l}
\textit{U}$_{f9}$%
\end{tabular}
&
\begin{tabular}{l}
$\frac{1}{2}\left( E^{(C^{4},C^{3},H^{2},H^{1},H^{3})}\otimes \sigma
_{z}^{(C^{1})}\otimes E^{(C^{2})}+E^{(C^{4},C^{3},H^{2},H^{1},H^{3})}\otimes
\sigma _{z}^{(C^{2},C^{2})}\right) $ \\
+$\frac{1}{2}\left( \sigma
_{z}^{(C^{4},C^{3},H^{2},H^{1},H^{3},C^{1})}\otimes E^{(C^{2})}-\sigma
_{z}^{(C^{4},C^{3},H^{2},H^{1},H^{3},C^{1},C^{2})}\right) $%
\end{tabular}
\\ \hline\hline
\end{tabular}
\newline

\newpage \textbf{Figure captions}\newline
FIG. 1. (a) A quantum network for realizing the seven-qubit entangling
transformation $U_{f9}$. Refocusing and decoupilng pulses are not included.
Vertical bars denote J-coupling gates, which are given by the unitary
transformation $exp\{-i\sigma _{z}^{(k)}\sigma _{z}^{(l)}\pi /4\}$ on spins
k and l. (b) A quantum circuit for implementing the seven-qubit modified D-J
algorithm. A function f is constant if and only if the output state for
every qubit is in all the \TEXTsymbol{\vert}0\TEXTsymbol{>} state, otherwise
a function is balanced. \newline
FIG. 2. The experimental spectra corresponding to the transformation ($%
U_{f1} $) for the constant function f1 from seven spins. Horizontal axes,
relative frequency (ppm); vertical axes, intensity (arbitrary units).\newline
FIG. 3. Same as Fig. 2, but (a), (b), (c), and (d) corresponding to the
balanced function transformations $U_{f2}$, $U_{f4}$, $U_{f5}$, and $U_{f9}$%
, respectively.

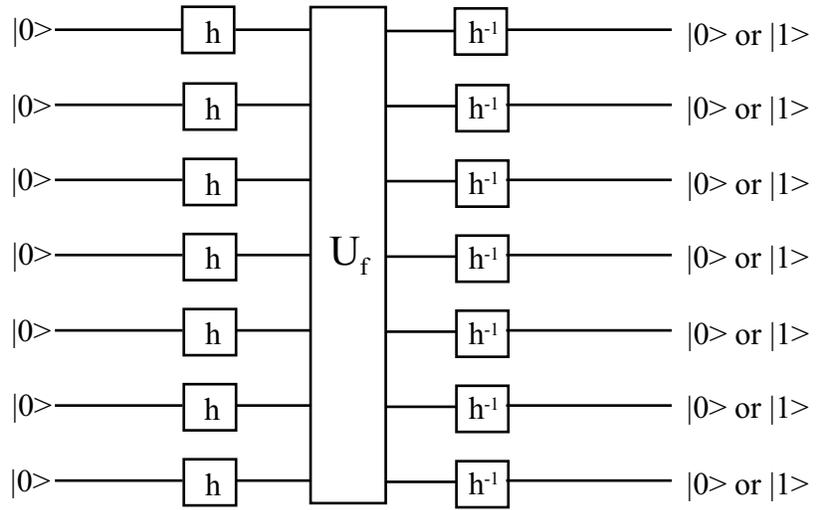
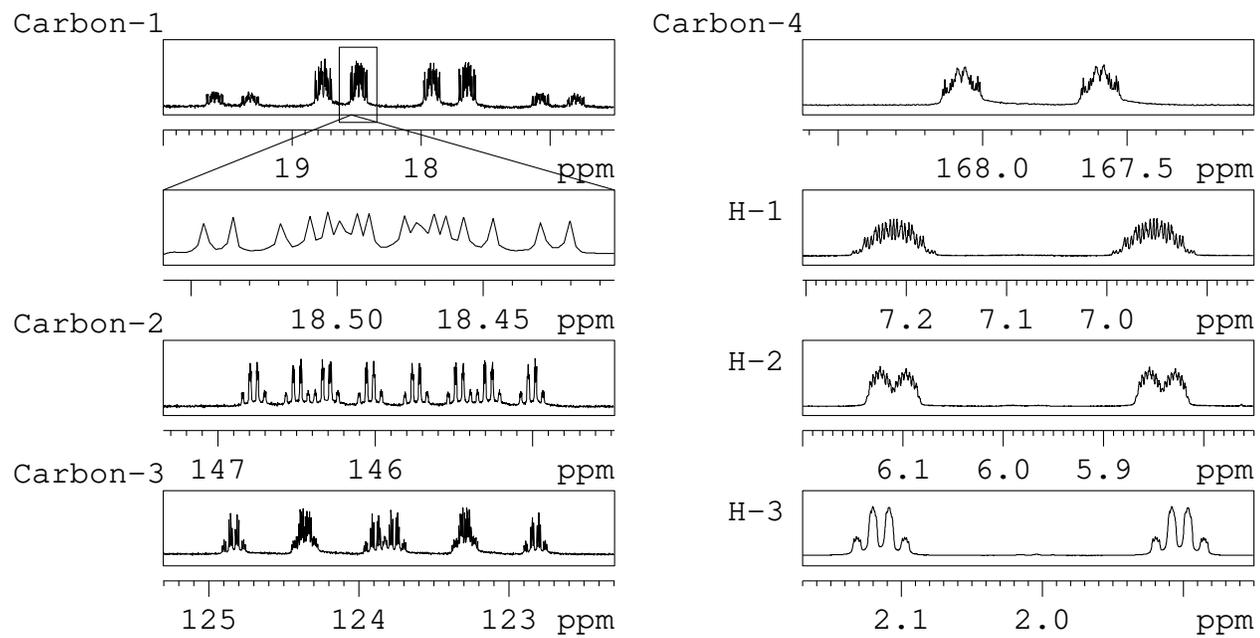
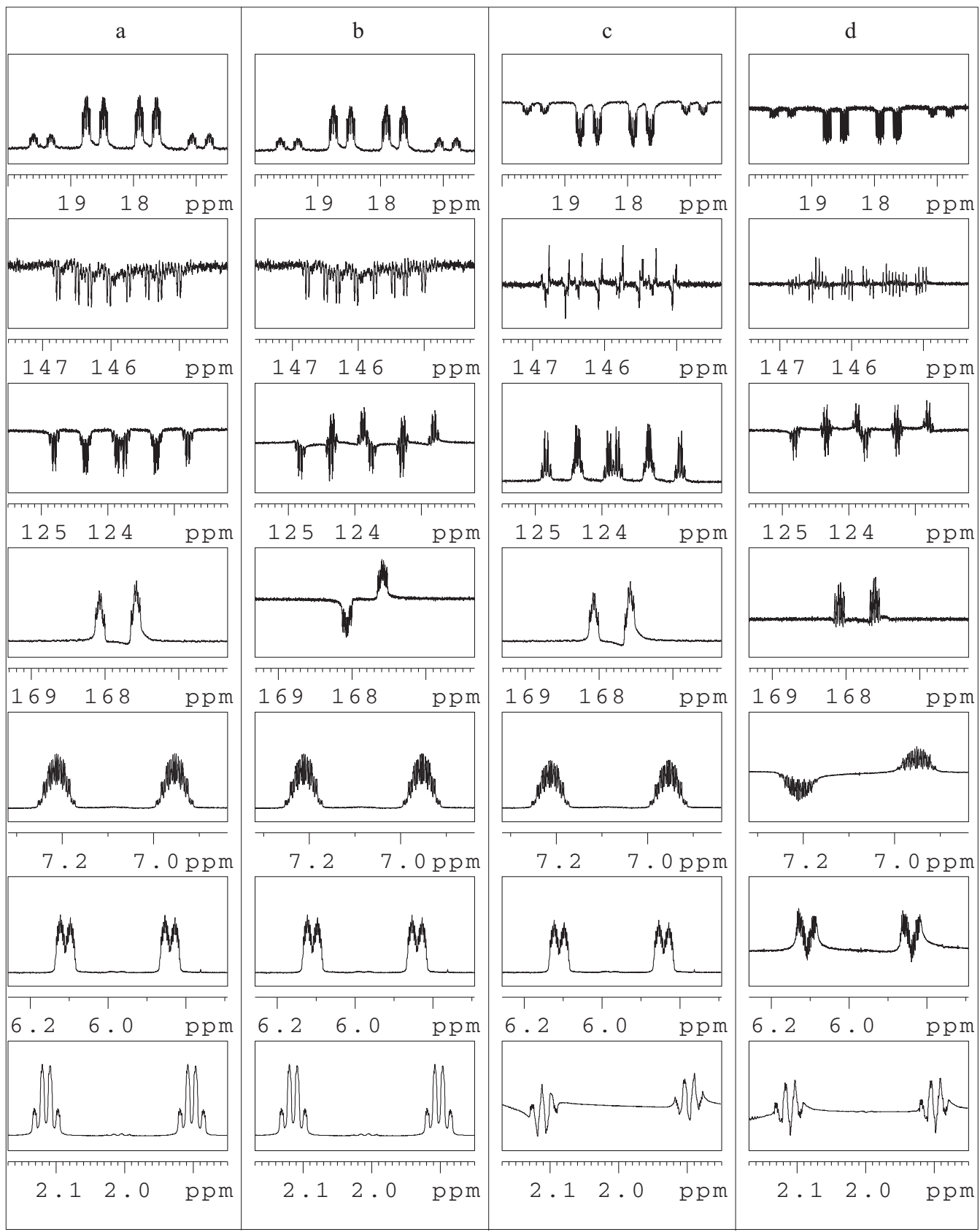
\end{document}